\documentstyle[preprint,prc,aps,eqsecnum,epsfig,amssymb]{revtex}
\tightenlines
\everymath={\displaystyle}
\begin{document}

\newcommand{\Sc}{Schr\"odinger}
\newcommand{\E}{equation}
 \newcommand{\al}{\alpha }
 \newcommand{\vfi}{\varphi }

\title{\bf Phase equivalent chains of
Darboux transformations in scattering theory}

\author{Boris F. Samsonov\footnote{e-mail : samsonov@phys.tsu.ru}}
\address{
Department of Quantum Field Theory, Tomsk State
University, 36 Lenin Ave., 634050 Tomsk, Russia}
\author{Fl. Stancu\footnote{e-mail: fstancu@ulg.ac.be}}
\address{University of  Li\`ege, Institute of Physics B5, Sart Tilman,
B-4000 Li\`ege~1, Belgium}

\date{\today}
\maketitle
\begin{center}
\begin{minipage}{0.99\textwidth}
\begin{abstract}
\baselineskip=0.50cm
\noindent

We propose a procedure based on phase equivalent chains of Darboux
transformations to generate local potentials satisfying the
radial Schr\"odinger equation and sharing the same
scattering data.
For potentials related by a chain of transformations
an analytic expression is derived for the Jost function.
It is shown how the same system of $S$-matrix poles can be
differently distributed between poles and zeros of a Jost function
which corresponds to different potentials with equal phase
shifts. The concept of shallow and deep phase equivalent potentials
is analyzed in connection with distinct
distributions of poles.
It is shown that phase equivalent chains do not violate the
Levinson theorem.
The method is applied to derive a shallow and a family of deep
phase equivalent
potentials describing the $^1S_0$ partial wave of the nucleon-nucleon
scattering.
\end{abstract}
\pacs {PACS numbers:}  03.65.Nk, 11.30.Pb, 21.45.+v, 13.75.Cs
\end{minipage}
\end{center}

\section{Introduction}
The construction of  potentials sharing the same scattering data,
called in the literature phase equivalent potentials, is one of the most
interesting applications of supersymmetric (SUSY) quantum mechanics
\cite{WI}
to the inverse scattering problem (see e.g. \cite{FADDEEV,Chadan-Sabatier}).
As it is known, the SUSY
approach,
when restricted to the derivation of new exactly solvable quantum
problems,
is basically equivalent to the
Darboux transformations method (see e.g.\cite{BAGSAM}).
Therefore we use SUSY and Darboux
transformations as synonyms.

To our knowledge, the simplest way of constructing phase
equivalent potentials
by using Darboux transformations, is due to Sukumar
\cite{SUKUMAR}.
It consists in a successive application of two
transformations with equal
factorization constants.
The method has been considerably developed,
up to an arbitrary modification of the discrete spectrum
\cite{BAYE}. It has been
extended to coupled channels \cite{SB97}
and used to derive deep and shallow phase equivalent
potentials describing
nucleon-nucleon \cite{SPAR97}
and nucleus-nucleus \cite{SPAR2000,SBL}
collisions. In these applications both deep and shallow potentials
have been constructed such as to share the same scattering data
at all energies. Shallow potentials possess only "physical"
states while the deep potentials, in addition to the
physical states also have "unphysical" bound states
which simulate the effect of the Pauli principle
(for details see e.g. \cite{SAITO,NEUD75,ACD91}).

Based on Darboux transformations, here we propose a new way
to derive deep and shallow phase equivalent potentials. Our aim
is to improve over the practical procedures of Refs.
\cite{SUKUMAR} and \cite{BAYE,SB97,SPAR97,SPAR2000,SBL}.
By replacing a chain of Darboux transformations
by an equivalent $N$th order transformation
and using a Wronskian
formula recently obtained in \cite{BFS95} (see also \cite{DASKALOV})
we use a {\it determinant} approach to construct
phase equivalent potentials.
When the {\it transformation functions} are sufficiently simple
(e.g. they are
solutions of the free particle Schr\"odinger equation)
this approach happens to be
more efficient
 than the one of previous works
\cite{SUKUMAR,BAYE,SB97,SPAR97,SPAR2000,SBL}.
The advantages of our method will be clearly seen in its
application to the nucleon-nucleon scattering potentials
considered here. In particular the potential tail will
be cured from the unwanted oscillations obtained in Ref. \cite{BAYE}.

As it is known (see e.g. \cite{SPAR97})
there are both deep and
shallow potentials corresponding to the same value of
the partial wave $l$.
The value of~ $l$ defines the long
distance behaviour of the potential \cite{FADDEEV}.
The short distance behaviour
\begin{equation}\label{short}
V(x)\longrightarrow \frac {\nu (\nu +1)}{x^2}\,,\quad x\to 0
\end{equation}
depends on the parameter $\nu $
which we shall call the {\it strength of the singularity at short
distances} or simply the {\it singularity strength}.
Following Ref. \cite{SPAR97}
we generally assume that $\nu $ is different from $l$.
In this work we consider only {\it $l$-preserving}
transformations and restrict
our discussion to the $l=0$ case,
although most of the derived formulae are valid in the general case.
In a subsequent paper \cite{SS} we shall consider {\it $l$-changing}
transformations which will allow to introduce higher
partial waves.

We assume that no Coulomb interaction is present in the
potential i. e. we are dealing with the Schr\"odinger equation
on the half axis, $x\in [0,\infty )$,
with the potential $V(x)$ satisfying the condition
(\ref{short}) and at long distances going to zero faster than
$x^{-2}$.
For every
eigenvalue $E$ the Schr\"odinger equation with such potentials
has two linearly independent solutions
with different behavior at short distances,
one
which behaves as $\sim x^\nu $, called {\it regular solution}, and
the other as $\sim x^{-\nu -1}$, called
{\it singular} (or {\it irregular}) {\it solution}.
In general Darboux transformations, which use
irregular solutions as transformation functions,
do not  preserve the property of a solution to remain
regular at the origin.
As a consequence, the change in the Jost function and the phase shift becomes
difficult to control since one has to know both the
Jost solution
of the initial equation
and
the derivative of this solution at the origin.
As far as we know
this problem has not been yet solved in the general context. But we have
found that there exist particular chains of transformations
where the derivative of the Jost solution of the initial equation
does not appear in the expression of the Jost function
for the transformed Hamiltonian. In this case one can get a closed
expression for the transformed Jost function in terms of
the initial Jost function and a rational function
of the momentum $k$ only.
Potentials having Jost functions related in such a way are known
in the literature as potentials of
the Bargmann class (see e.g. \cite{FADDEEV}).

Once we know how the Jost function is changed by
a chain of Darboux transformations we can select chains
that produce potentials with the same $S$-matrix and hence
with the same phase shift.
Such potentials are called in the literature
{\sl phase equivalent}
although a more appropriate term could be {\sl isophase} potentials since
they give exactly the same phase shift
at all energies. Here we shall use both terms.
So, to get phase equivalent potentials we propose
to use {\sl phase equivalent chains}
instead of an integral form of
Darboux transformations used in
\cite{SUKUMAR,BAYE,SB97,SPAR97,SPAR2000,SBL}.
We start from the same exactly solvable reference potential
and the action of each chain will
result in potentials always having the same $S$-matrix.
Finally we note that
this approach is basically
equivalent to the inverse scattering method where the kernel of
the Fredholm operator involved in the Gelfand-Levitan-Marchenko
equation is degenerate \cite{FADDEEV} or to
a supersymmetric transformation
\cite{SUKUMAR,BAYE,SB97,SPAR97,SPAR2000,SBL},
but it is more direct and has more computational advantages.

The paper is organized as follows. In the next section we recall the Darboux
transformation method in the context of scattering theory.
In Sec. III we
introduce special chains for which a closed expression for the
transformed Jost function is derived. Then we select
 chains which lead to families of phase equivalent
 (or isophase) potentials.
Sec. IV is devoted to an application of our method to the derivation of deep
and shallow phase equivalent potentials for the nucleon-nucleon
scattering.
Conclusions are drawn in the last section.
Details of important analytic calculations are gathered in
the Appendix.

\section{Darboux transformation method}

The method of Darboux transformations, well known in soliton
theory \cite{matveev}, finds now more and more
applications in other fields of theoretical and mathematical
physics \cite{SUSY}.
In this section we recall the definition of
Darboux transformations and
give their main properties necessary in subsequent sections.

Essentially, the method consists in getting
solutions $\vfi$ of one Schr\"odinger equation
\begin{equation}\label{trans}
h_1\vfi =E\vfi ,\quad h_1=-\frac{d^2}{dx^2}+V_1(x)~,
\end{equation}
when solutions $ \psi$ of another equation
\begin{equation}\label{init}
h_0\psi =E\psi ,\quad h_0=-\frac{d^2}{dx^2}+V_0(x)~,
\end{equation}
are known. This is achieved by acting with a differential operator
$L$
of the form
\begin{equation}\label{L1}
\vfi =L\psi , \quad L=-d/dx +w(x)\,,
\end{equation}
where the real function $w(x)$, called {\it superpotential}, is defined as the
logarithmic derivative of a known solution of (\ref{init}) denoted by $u$
in the following. One has
\begin{equation}\label{SUPER}
w=u'(x)/u(x)\,,\quad h_0u=\al u\,,
\end{equation}
with $\al \le E_0$, where
$E_0$ is the ground state energy of $h_0$ if it has a discrete
spectrum or the lower bound of the continuum spectrum otherwise.
The function $u$ is called {\it transformation}
or {\it factorization function} and $\al $ its {\it factorization
constant} or {\it factorization energy}. The potential $V_1$ is defined
in terms of the superpotential $w$ as
\begin{equation}\label{V1general}
V_1(x)=V_0(x)-2w'(x)\,.
\end{equation}

The operators $L$ and $L^+=d/dx +w(x)$ establish a one-to-one
correspondence between two dimensional subspaces of solutions of
Eqs. (\ref{trans}) and (\ref{init}) with a given
$E\ne \al $. When $E=\al $ the transformed equation  (\ref{trans}) has
$v=1/u$ and $\tilde v$
as linearly independent solutions. The latter
can be found by using
the Wronskian property $W(v,\tilde v) = 1$
which gives
\begin{equation}\label{tv}
\tilde v=v\int \limits _{x_0}^xv^{-2}(x)dx~,
\end{equation}
where $x_0$ is an arbitrary constant.
Hence,
the knowledge of all solutions of the initial equation provides
the knowledge of all solutions of the transformed equation and
in particular of all physical solutions
in the usual quantum mechanical sense.

Since the above procedure does not depend on a particular choice
of the potential $V_0$ the transformed Hamiltonian $h_1$ can play
the role of the initial Hamiltonian for the next transformation
step. In such a way one gets a chain of exactly solvable
Hamiltonians $h_0,\ h_1,\ \ldots \ h_N$ with the potentials $V_0,\ V_1,\
\ldots\ V_N$.
To avoid any confusion we mention that everywhere,
 except for especially mentioned cases,
we shall use subscripts to
distinguish between quantities related to different Hamiltonians,
$h_0$, $h_1$,\ldots and shall omit them when discussing general
properties regarding all Hamiltonians.

A first order Darboux transformation operator $L_{j\,,\,j+1}$
as defined by (\ref{L1})
relates two Hamiltonians $h_j$ and $h_{j+1}$. If one is not
interested in the intermediate Hamiltonians $h_1$,~\ldots ,~$h_{N-1}$
and all factorization energies are chosen to be different
from each other the
whole chain may be replaced by a single
transformation
given by an $N$th order transformation operator
denoted by $L^{(N)}$, defined as a succession of $N$ first
order transformations.
A compact representation of this operator is given by
\cite{CRUM}
 \begin{equation}\label{psiN}
\psi _N (x,k)=L^{(N)}\psi _0(x,k)=
W(u_1,\ldots ,u_N,\psi _0(x,k))\,W^{-1}(u_1,\ldots ,u_N)~,
 \end{equation}
where $\psi _0(x,k)$ is a solution of Eq. (\ref{init})
corresponding to the energy $E=k^2$
and $\psi _N (x,k)$
satisfies
\begin{equation}\label{ORDERN}
h_N\psi _N (x,k)=E\psi _N (x,k)\,,\quad E=k^2~.
\end{equation}
The transformation functions, although labeled by a subscript,
are eigenfunctions of the initial Hamiltonian
\begin{equation}
h_0 u_j(x)=\al _ju_j(x)\,,\quad \al_j=-a_j^2~.
\end{equation}
They should be chosen such as the Wronskian
$W(u_1,\ldots ,u_N)$ is nodeless in the interval $(0,\infty )$.
This condition guarantees the absence of singularities in the
potential
\begin{equation}\label{VN}
V_N=V_0-2\frac {d^2}{dx^2}\log W(u_1,\ldots ,u_N)~,
\end{equation}
defining the Hamiltonian $h_N$
of (\ref{ORDERN}),
inside this interval. The formula (\ref{psiN}) is valid for
any $E=k^2$ except for $k=ia_j$ ($j=1,\ldots N$).
For these values
of $k$ the corresponding solutions are
\begin{equation}\label{psiNj}
\psi_N(x,ia_j)=
W^{(j\,)}(u_1,\ldots ,u_N)W^{-1}(u_1,\ldots ,u_N),\quad
j=1,\ldots ,N~,
\end{equation}
where $W^{(j\,)}(u_1,\ldots ,u_N)$ is the $(N-1)$th order Wronskian
constructed from $u_1,\ldots ,u_N$ except for
$u_j$.

In general, a chain of $N$ transformations can have a number of
elements corresponding to transformation functions
with identical factorization constants. In the simplest $N = 2$
case where the first transformation is implemented by the
function $u_1=u$ and the second by
$v=1/u$ one gets back the initial potential $V_0$. This means that
the operator $L^+$ realizes the transformation in the opposite
direction. More generally, the second
transformation can be made with
the linear combination
$v_1=cv+\tilde v$ where $\tilde v$ is given by (\ref{tv}). The
action of such a chain leads to an eigenfunction of $h_2$ of the form
\begin{equation}\label{INTEGRAL}
\psi _2(x,k)=\psi _0(x,k)-\frac{u(x)}{W_2(x)}
\int\limits _{x_0}^xu(t)\psi_0(t,k)dt,
\end{equation}
where
\begin{equation}\label{W2}
W_2(x)= c+\int\limits_{x_0}^x u^2(t)dt~.
\end{equation}
Choosing  $x_0=0$ (or
$x_0=\infty $) together with the condition $c\ge 0$
(or $c\le 0$) guarantees a nodeless
$W_2(x)$ inside the interval $(0,\infty )$.
The integral representation (\ref{INTEGRAL}) based on two
subsequent transformations with the same factorization
constants was used in Refs.
\cite{SUKUMAR,BAYE,SB97,SPAR97,SPAR2000,SBL}
for getting phase equivalent potentials. We now describe an
alternative approach which has the advantage that
the above $N = 2$ chain of transformations
 can
be included in a larger chain as a subchain.

The potential $V_2(x)$ is a particular case of
(\ref{VN}) where $W(u_1,\ldots ,u_N)$ is given by
$W_2(x)$ of (\ref{W2}).
It has been
shown \cite{BFS95} that the function $W_2$ is nothing else but the
limiting value of the ratio $W(u_1,u_2)/(\al_1-\al_2)$
when $u_2\to u_1$. This fact, together with a property of the Wronskians
established in \cite{BFS95,DASKALOV},
which allows to reduce the order of the
determinants involved in calculations,
implies that one can use the same formulae (\ref{psiN}) and
(\ref{VN}) also in the case of pairwise identical factorization
constants but the Wronskians should be calculated with the help of
another representation.
If the Wronskian in
(\ref{psiN}) is of even order, $N=2m$, it
becomes proportional to an $m\times m$ determinant
\begin{equation}\label{W2m}
W(u_1,\ldots u_{2m})=
\prod\limits_{l\ne j}^{2m}(\al_l-\al_j)\times
\left|
\begin{array}{ccccc}
W_{\al_1,\,\, \al_2} &W_{\al_1,\,\,\al_4}&W_{\al_1,\,\,\al_6}
&\cdots &W_{\al_1,\,\,\al_{N}}\\
W_{\al_3,\,\, \al_2} &W_{\al_3,\,\,\al_4}&W_{\al_3,\,\,\al_6}
&\cdots &W_{\al_3,\,\,\al_{N}}\\
W_{\al_5,\,\, \al_2} &W_{\al_5,\,\,\al_4}&W_{\al_5,\,\,\al_6}
&\cdots &W_{\al_5,\,\,\al_{N}}\\
\vdots &  \vdots &  \vdots &  \ddots &  \vdots \\
W_{\al_{N-1},\,\, \al_2} &W_{\al_{N-1},\,\,\al_4}&W_{\al_{N-1},\,\,\al_6}
&\cdots &W_{\al_{N-1},\,\,\al_{N}}
\end{array}
\right|~.
\end{equation}
For an odd order Wronskian, $N = 2m + 1$, one has
$$
\begin{array}{l}
W(\psi ,u_1,\ldots u_{2m})=
\prod\limits_{l\ne j}^{2m\vphantom{M_{M_k}}}(\al_l-\al_j)
\prod\limits_{k=1}^{m\vphantom{M_{M_k}}}(\al_{2k}-E)\\
\hphantom{WWWWWWWWWWWWWWWWWWWWWWWWWWWWWWWWWWWWWWWWWWWWWWWW}
\end{array}\vspace{-3em}
$$
\begin{equation}\label{W2m1}
\begin{array}{l}
\hphantom{WWWWWWWWWWWWWWWW}\\
\times
\left|
\begin{array}{cccccc}
\psi &W_{E,\,\, \al_2} &W_{E,\,\,\al_4}&W_{E,\,\,\al_6}
&\cdots &W_{E,\,\,\al_{N}}\\
u_1 &W_{\al_1,\,\, \al_2} &W_{\al_1,\,\,\al_4}&W_{\al_1,\,\,\al_6}
&\cdots &W_{\al_1,\,\,\al_{N}}\\
u_3 &W_{\al_3,\,\, \al_2} &W_{\al_3,\,\,\al_4}&W_{\al_3,\,\,\al_6}
&\cdots &W_{\al_3,\,\,\al_{N}}\\
u_5&W_{\al_5,\,\, \al_2} &W_{\al_5,\,\,\al_4}&W_{\al_5,\,\,\al_6}
&\cdots &W_{\al_5,\,\,\al_{N}}\\
\vdots &  \vdots &  \vdots &  \vdots &  \ddots &\vdots \\
u_{N-1}&W_{\al_{N-1},\,\, \al_2} &W_{\al_{N-1},\,\,\al_4}&W_{\al_{N-1},\,\,\al_6}
&\cdots &W_{\al_{N-1},\,\,\al_{N}}
\end{array}
\right|~,
\end{array}
\end{equation}
where
\begin{equation}
W_{\al_k,\,\,\al\,_l}=\frac{W(u_{\,k},u_{\,l})}{\al_k-\al_l}\,,\quad
W_{E,\,\,\al\,_l}=\frac{W(\psi ,u_{\,l})}{E-\al_l}\,.
\end{equation}
The product $\prod_{l\ne j}^{2m}(\al_l-\al_j)$
introduced above contains
only the factors $(\al_l-\al_j)$ appearing in
$W_{\al\,_k,\,\,\al\,_l}$.
For identical factorization constants the function $W_{\al_k,\,\,\al\,_l}$
has to be calculated from the formula (\ref{W2}).

\section{PHASE EQUIVALENT CHAINS OF TRANSFORMATIONS}

Our method is based on a choice of
transformation functions which allows to
transform the Jost function through a special chain of
Darboux transformations. We follow the usual definitions of the
scattering theory which can be found for example in Refs.
\cite{FADDEEV,Chadan-Sabatier}.

\subsection{Jost function for a special chain of transformations}

Let us
choose the following set of eigenfunctions of the Hamiltonian $h_0$
\begin{equation}\label{uv}
v_{1}(x)\,,\ldots \,, v_\nu (x)\,,
u_{\nu +1}(x)\,,  v_{\nu +1}(x)\,,\ldots \,, u_n(x)\,, v_n(x)\,,
\quad \nu\ge 0~,
\end{equation}
\begin{equation}
h_0u_j(x)=-a^2_ju_j(x)\,,\quad h_0v_j(x)=-b^2_jv_j(x)\,,\quad
{\rm Im}\,a_j={\rm Im}\,b_j=0~,
\end{equation}
as transformation functions for the Darboux
transformation of order $N=2n+\nu $.
The role of the first $\nu $ functions $v_j(x)$ which are absent
in (\ref{uv}) for $\nu =0$ will be clear below.
Let the $a$'s and the $b$'s be different
from each other and choose them such as the corresponding factorization
energies be smaller than the ground state energy of $h_0$ when it
has a discrete spectrum or less than the lower bound of the
continuous spectrum otherwise.

We distinguish between the functions $u$ and $v$ from their behaviour at
the origin.
The functions $v$ are regular, $v_j(0)=0$,
 and hence
are uniquely defined up to a constant factor which is not essential for
our purpose.
They form a {\it regular
family}
and are exponentially increasing  at
large $x$ and therefore it is natural to restrict all $b$'s to
be positive.
The functions $u_j$
are irregular at the origin, $u_j(0)\ne 0$, and from a {\it singular family}.
They are defined as linear combinations of two Jost
solutions $f(x,\pm k)$:
\begin{equation}\label{uj}
u_j(x)=A_jf(x,-ia_j)+B_jf(x,ia_j)~.
\end{equation}
When $A_j\ne 0$ the functions $u_j$ increase exponentially at large $x$
provided $a_j>0$. At fixed $a_j$ they form a one-parameter ($A_j/B_j$)
family.
The case $A_j=0$, $B_j\ne 0$, with $a_j>0$ is equivalent to the case
$A_j\ne 0$, $B_j=0$, with $a_j<0$. They correspond to a
decreasing exponential function (\ref{uj}),
uniquely defined (up to a constant factor).
In this case in order to stress the
asymptotical decrease of this function
we allow $a_j$ to be negative and set,
$B_j=0$, $A_j\ne 0$.
So, for $a_j>0$ the functions (\ref{uj}) have an increasing
asymptotic behavior and
for $a_j<0$ their
asymptotic behavior is opposite.
It is useful to note that if $A_j/B_j=F_0(ia_j)/F_0(-ia_j)$
the function (\ref{uj}) becomes regular. In the chain of
functions (\ref{uv}) we shall allow to move some functions from
the singular family to the regular family. This preserves the
number of functions $N$ but changes the value of $\nu $. The
effect of such a modification of a chain will be discussed
below.
The particular case where the initial potential is zero and
$A_j=B_j$ has been considered earlier in Ref. \cite{SamShek}.

In the Appendix we show that
the chain of transformations with the transformation functions
(\ref{uv}) increase the singularity parameter of the initial
potential
by $\nu $ units, $\nu_0\to \nu _N=\nu_0+\nu$,
and the transformed Jost function
$F_N(k)$ is related to the initial function $F_0(k)$ by
\begin{equation}\label{MNnu}
F_N(k)=
F_0(k)\prod\limits_{j\,=\,1}^{\nu }\frac{k}{k+ib_j}
\prod\limits_{j\,=\,\nu+1 }^n\frac{k-ia_j}{k+ib_j}~.
\end{equation}
Since a Jost function is analytic in the upper half of
the complex $k$-plane
(see \cite{FADDEEV,Chadan-Sabatier})
all $b$'s must be positive whereas the $a$'s
can have any sign,
which agrees with our sign convention for $a_j$ and $b_j$ and
every positive $a_j$ corresponds to a
discrete level $E_j=-a_j^2\,$ of $V_N$.
A remarkable
property of the relation (\ref{MNnu}) is that it is independent of
the ratio $A_{j}/B_j$ with $A_{j}$ and $B_j$ from the definition (\ref{uj}).
But the potential $V_N$ does
depend on this ratio.
Hence, keeping $F_N(k)$ fixed but varying $A_{j}/B_j$
such as $u_j$ remains irregular
we can obtain
a family of isospectral potentials.
Nevertheless
the potential $V_N$ belongs
to this isospectral family only if
$A_j\ne 0$
and the function $u_j(x)$ increases at large $x$.
The choice $A_j=0$ is equivalent to
changing the sign of $a_j$ and leads to
losing the level $E=-a_j^2$ in the potential $V_N$.
This property does not have an analogue in the usual scheme of
SUSY transformations \cite{SUSY}.

\subsection{Phase equivalent chains}

Now we turn to the construction of phase equivalent or
{\sl isophase} potentials.
Such potentials have identical $S$-matrices which are
defined in terms of the Jost function by
\begin{equation}\label{Sk}
S(k)=e^{2i\delta (k)}=\frac{F(-k)}{F(k)}~.
\end{equation}
Therefore if for a real $k$ two Jost
functions differ from each other by a real factor they
correspond to isophase potentials. In particular, these potentials can have
equal Jost functions. Taking into consideration the discussion
at the end of the previous subsection
we see that by changing the ratio $A_j/B_j$, which
corresponds to a discrete level of the potential $V_N$,
we can get not only isospectral but also isophase potentials.
As it is clear from the Appendix, the potentials are isospectral
provided the function $u_j$ remains singular.
If it moves from the
singular to the regular family, the transformed potential
does not belong to the family of isospectral potentials any more
but as we show below it remains in the family of isophase
potentials.

Let us suppose for example that the function $u_1(x)$
becomes regular from being singular. Then for a regular $u_1$ the Jost
function of the potential $V_N(x)$ corresponds to (\ref{MNnu})
with $\nu =0$ and for a singular $u_1$ it should be calculated
by the same formula where $\nu =2$ and $b_1$ is replaced by
$a_1$. From here it follows that the ratio of these two Jost
functions is equal to $k^2/(k^2+a_1^2)$ which is real
for a real $k$
and hence these two chains of transformations are isophase. It is also
clear that if $\nu $ functions from the singular family move
to the regular family
(i.e. $\nu $ zeros of the Jost function are transformed into its
poles)
 the potential $V_N$
loses $\nu $ discrete levels, and acquires an additional
singularity strength equal to $2\nu $, but remains in the family of
isophase potentials.
Furthermore, this procedure is completely reversible. If one has a
singular potential with $\nu\ge 2$
then one of the
poles of its Jost function can be transformed into a zero
in the upper half of the complex $k$-plane
which leads to the creation of a discrete level and a decrease of the
singularity strength $\nu \to \nu -2$.
This procedure can be repeated
either  up to the point where
the potential loses all
discrete levels and all free parameters,
becoming uniquely defined by its phase shift
(see e.g. \cite{FADDEEV,Chadan-Sabatier})
or up to the point where $\nu$ becomes equal to zero or one.
To avoid
additional singularities inside the interval $(0,\infty )$
one has to take these poles such as the additional level be
located below the existing discrete levels. The latter
restriction is well known in the SUSY approach \cite{SUSY}.

The above method of deriving isophase potentials is not unique.
As it is known one can create a new energy level independent
of the existence of poles of the Jost function,
at any desirable position even in the middle of the continuous
spectrum,
 and preserving
the $S$-matrix.
For this purpose one can use a two-step Darboux
transformation with the same factorization
energy
\cite{SUKUMAR,BAYE,SB97,SPAR97}.
After a new level has been created the
singularity of the potential loses two units
$\nu \to \nu -2 $ \cite{BAYE}.
This procedure can be repeated up to a limit where the
singularity parameter becomes equal to either zero or one.
For a regular potential or a potential with $\nu =1$ no
possibility exists to create a level by a phase preserving
procedure.
In this context the smallest possible value is $\nu =2 $
when the potential with an additional
discrete level becomes regular after transformation.
According to Ref. \cite{BFS95}
a chain of two transformations with coinciding factorization
constants
is equivalent to considering the
limit $a_{j+1}\to -\,a_j$ and using the Wronskian
formulae (\ref{W2m}) and ({\ref{W2m1}).
Without loosing generality
let us consider $a_2\to -a_1$ for our system of
transformation functions (\ref{uv}) with $\nu =0$.
In this limit
the Jost function (\ref{MNnu}) where we set $\nu =0$
\begin{equation}
F_N(k)=
F_0(k)\frac{k^2+a_1^2}{(k+ib_1)(k+ib_2)}
\prod\limits_{j\,=\,3}^n\frac{k-ia_j}{k+ib_j}~.
\end{equation}
has no singularity.
The corresponding potential $V_N$ can be considered as a result
of the application
of the above two-step phase preserving SUSY transformation on a
singular potential $V_{N-2}$
generated by the system
of transformation functions
$v_1,v_2,u_3,v_3,\ldots, u_n,v_n$
whose Jost function $F_{N-2}$ is given by the same
formula (\ref{MNnu}) with $\nu =2$.
Finally we conclude that $V_{N-2}$ is isophase with $V_N$
since the ratio $F_N/F_{N-2}=(k^2+a_1^2)/k^2$ is real.

\subsection{The $S$-matrix poles}

Once the Jost function $F(k)$ is known one can
calculate the phase shift $\delta (k)$ by using the definition
(\ref{Sk}).
The asymptotic behavior of the scattering wave
function of the potential $V_N$
with $l=0$ and singularity strength $\nu $ at long distances is
\begin{equation}\label{psilong}
\psi_N (x,k)\sim \sin (kx -\textstyle{\frac 12}\nu \pi+\delta_N (k)),
\quad x\to \infty~.
\end{equation}
In our approach $F_N(k)$
differs from $F_0(k)$ by
a rational function of momentum $k$ and
therefore the expression for $\delta_N (k)$ becomes rather complicated when the
number of transformation functions is sufficiently large.
An equivalent expression which is more
convenient for practical  calculations is
\begin{equation}\label{deltaN}
\delta_N(k)=\delta_0(k)-\sum\limits_{j\,=\,{\nu +1}}^n\,\arctan (k/a_j)
-\sum\limits_{j\,=\,{1}}^n\,\arctan (k/b_j)~,
\end{equation}
in agreement with the result obtained in \cite{BAYE}.
 From (\ref{deltaN}) it follows that the
$S$-matrix acquires additional poles at  $a_j$ and
$b_j$ through the application of the
transformation $L^{(N)}$ consistent with the discussion
of the previous section.

For $l=0$ the
effective range expansion is (see e. g. \cite{BJ})
\begin{equation}\label{range}
\lim \limits_{k\to 0} k\cot \delta(k)=
-\frac 1a+\frac 12 r_0k^2-Pr_0^3k^4+\ldots\,~,
\end{equation}
where $a$ is the scattering length and $r_0$
effective range. Expanding the phase shift (\ref{deltaN}) in a power series
one obtains
\begin{equation}\label{aN}
\frac{1}{a^{(N)}}=\frac{1}{a^{(0)}}+
\left[\,\sum\limits_{j\,=\,{\nu +1}}^na_j^{-1}+
\sum\limits_{j\,=1}^nb_j^{-1}\right]^{-1}~,
\end{equation}
and
\begin{equation}\label{rN}
r^{(N)}=r^{(0)}+
\frac {2a^{(N)}}{3} \left[1-
{a^{(N)}}^{-3}
\left(\,\sum\limits_{j\,=\,\nu +1}^n a_j^{-3}+
\sum\limits_{j\,=1}^nb_j^{-3}\right) \right]~.
\end{equation}
These formulae can be used to calculate theoretical values
to be compared with experimental results, as for
example those obtained in the neutron-proton scattering experiments.
But first one has to fit the parameters $a_j$ and $b_j$ in (\ref{deltaN})
such as to reproduce the phase shift data as close as possible.
As a result of the fit one gets a sequence of  $n_+$ positive and
$n_-$ negative poles of the $S$-matrix
all having absolute values below the pole
associated with the ground state of $h_0$ if this
has a discrete spectrum or below the lower bound of its
continuum spectrum otherwise. (We shall comment below on the
cases when this is not so.)
Every negative pole is associated
with a singular transformation function decreasing at infinity
and belonging to the singular family of (\ref{uv}).
This is just the Jost solution of the initial equation
at $k_j=ia_j$ which is uniquely defined.
A positive pole corresponds to an increasing transformation
function (\ref{uj}), which can be either regular or
singular at $x=0$. When regular it is uniquely
defined, whereas  singular functions
form a one-parameter family.

According to the choice of transformation functions (\ref{uv}) our
method is restricted by the condition $n_-\le n_+$ and one has
to distribute the functions related to positive poles among
regular and singular families such as the number of
singular transformation functions does not exceed the number of
regular ones. This means that at least $n_-$ functions related
with positive poles should be regular.
For the remaining positive poles there are different possibilities
either related to the regular $\left\{v_j\right\}$ family or
to the singular $\left\{u_j\right\}$ family. All these possibilities
correspond to the same phase shift and hence will give different phase
equivalent chains of transformations.
When all the remaining positive poles are
associated with the regular family
the potential will acquire the
largest singularity strength $\nu =n_+-n_-$ and according to the
formula (\ref{MNnu})
will have no additional discrete levels
with respect to $V_0$.
In this case every function from the system (\ref{uv}) is
uniquely defined and the potential $V_N$ has no free parameters.
This agrees with the well-known statement of the inverse
scattering method
(see e.g. \cite{FADDEEV,Chadan-Sabatier}) that in the absence
of discrete spectrum a
local potential is uniquely
defined by its phase shift.
Such potentials are called in the literature
{\sl shallow potentials}.

When a maximal number of positive poles (equal to
$\gamma =(n_+ - n_-)/2$ for $N$  even and
$\gamma =(n_+ -n_- -1)/2$ for $N$ odd)
is included in the singular family every such pole
corresponds to
an additional
discrete level of $V_N$ and this potential
will acquire the smallest singularity strength
$\nu =0$ (regular potential) for $N$ even and $\nu =1$ for $N$ odd
and $\gamma $ additional discrete levels.
Every positive pole $a_j$ associated with a singular function
$u_j$ will correspond to a discrete level $E_j=-a_j^2$ of $V_N$.
According to (\ref{uj}) the function $u_j$ has a free parameter
$A_j/B_j\,$. Thus we obtain a $\gamma $-parameter
family of isophase and isospectral potentials.
Such potentials are called in the literature
{\sl deep potentials}. Of course there are also intermediate possibilities.
When $n_+=n_-$
or $n_+=n_-+1$
one can construct only a
deep (regular if $n_+=n_-$) potential
without additional discrete levels. For this potential
phase preserving transformations which at the same
time do not change the angular momentum $l$ are impossible.

When $\nu =n_+-n_-\ge 2$ one can construct a $2\gamma $-parameter
family of isophase potentials by introducing
$\gamma $  energy levels into the
shallow potential.
For this purpose one has to use a chain of
$N+2\gamma $ transformations, phase equivalent to the initial chain
of $N$ transformations, to which $2\gamma $
transformation functions with pairwise equal factorization constants
$a_j^2 = b_j^2$
should be added. To calculate the Wronskian of such a system
one has to
use either the formula (\ref{W2m}) or (\ref{W2m1}). The potentials
$V_N$ form a $2\gamma $-parameter family where $\gamma $
energy levels play  the role of parameters. The other $\gamma $
parameters correspond to the constant $c$ of
(\ref{W2}).

It may happen that the absolute values of some poles is
above the position defined by an excited state energy of $h_0$.
Then using a phase preserving procedure,
we first have
to delete all such levels including ground state level
from the spectrum of $h_0$
and then use
the above described technique.
This means to include
additional pairs of functions with equal factorization constants
in the system of transformation functions (\ref{uv}).
The corresponding energies have to
coincide with the positions of the discrete levels to be deleted,
which evidently does not change the fitted phase shift.
The Wronskian of the entire system of
transformation functions has to be calculated by using either
(\ref{W2m}) or (\ref{W2m1}).

When the number of negative poles exceeds the number of
positive ones, the problem becomes more sophisticated.
In some cases it can be solved by introducing
transformations changing
 the orbital angular momentum $l$.
We are planning to discuss these problems in a subsequent paper \cite{SS}.

In this subsection all potentials have an
exponential decrease at infinity, provided the initial
potential corresponds to $l=0$. Therefore all transformed
potentials also correspond to $l=0$.
An useful remark is that when fitting an experimental phase shift
 one can consider either real
poles of the $S$-matrix  or pairwise mutually conjugated
poles. In either case they should be associated with the regular
family of the system (\ref{uv}). The corresponding Wronskian
is either real or purely
imaginary, since the transformation functions are
mutually conjugated. The potential difference
$V_N - V_0$ resulting from
(\ref{VN}) is real in either case.

\subsection{The Levinson theorem}

Here we show that the formula (\ref{deltaN})
is in agreement with the Levinson theorem in its generalized
form \cite{SWAN} (see also
\cite{SPAR97}).
For a potential satisfying the condition (\ref{short})
and decreasing faster than $x^{-2}$ at long distances
this theorem states that
\begin{equation}\label{Levinson}
\delta (0)-\delta (\infty )=
[{\cal N}+\textstyle{\frac 12}(\nu -l)]\pi~,
\end{equation}
where ${\cal N}$ is the number of bound states
$l$ the partial wave and $\nu$ the singularity strength.
In our case we have to set $l=0$.
Suppose that this theorem holds for the initial potential $V_0$.
We shall show that in this case it also holds for the
transformed potential $V_N$. Since the right hand side of
(\ref{Levinson}) is a linear form in the variables
${\cal N}$ and  $\nu $ it remains to show
that  this formula is valid for the case where
$\delta (0)$, $\delta (\infty )$, ${\cal N}$ and  $\nu $
represent differences between the same quantities associated with the
potentials $V_0$ and $V_N$.
In what follows
in this subsection,
by these quantities we understand the corresponding increments.

The potential $V_N$ is supposed to be obtained from $V_0$ by an
$l$-preserving chain of transformations with transformation
functions (\ref{uv}).
When all additional positive poles of the $S$-matrix are
related to the regular family, the potential $V_N$ has the same
number of bound states as $V_0$ but its singularity strength
is $\nu_N=\nu =n_+-n_-$\,. Using (\ref{deltaN}) one finds
that $\delta (0)=0$, $\delta (\infty )=-\pi (n_+-n_-)/2$ which
is in the agreement with (\ref{Levinson}) at ${\cal N}=0$. If
now we transform ${\cal N}\le (n_+-n_-)/2$ regular functions into
singular ones the potential $V_N$ acquires ${\cal N}$ additional
discrete levels and loses $2{\cal N}$ units of singularity strength
which also is in agreement with the relation (\ref{Levinson}).
Hence, we conclude that $l$-preserving chains of
transformations with transformation functions $(\ref{uv})$ do not
violate the Levinson theorem.

\section{Application to neutron-proton elastic scattering}

Here we apply our method to derive shallow and deep
phase equivalent potentials, both reproducing the
neutron-proton ($n\,p$) experimental elastic
scattering phase shift for the $^1 S_0$ partial wave in the
laboratory energy interval 0-350 MeV.
The experimental values taken from Ref. \cite{STOKS} 
are displayed
in Table I.
Our choice
$V_0(x)=0$ which allows to use
solutions of the free particle Schr\"odinger equation simplifies the
calculations considerably.
The theoretical phase shift also displayed in Table I under $\delta_6$ 
is obtained from the formula (\ref{deltaN})
with $\delta_0(k)=0$ and $\nu =0$ where we had to fix $n$
optimallly and search
for $a_j$
and $b_j$ ($j$ = 1,2,...$n$).  By using
the standard Mathematica package NonlinearRegress \cite{MATHEMATICA}
we found that we need six $S$-matrix poles
in order to fit the experimental phase shift very closely.
Thus we have $N=2n=6$ and the poles are
\begin{equation}\label{1s0poles}
\begin{array}{l}
a_1=-0.0401 ,\  \ a_2= -0.7540 ,\ \ a_3= 4.1650 ,\ \ \\
b_1=\ \ \ 0.6152 ,\ \  b_2=\ \ \  2.0424 ,\ \ b_3=4.6000,
\end{array}
\end{equation}
in fm$^{-1}$ units.
In the notation of Subsection III C we have $n_-=2$ and $n_+=4$. From
Eqs. (\ref{aN}) and (\ref{rN}) with $1/a^{(0)}=0$ and
$r^{(0)}=0$ we obtain $a^{(6)}=-23.7032$ fm and $r^{(6)}=2.6235$ fm.
The comparison
with the recommended values $a_{exp}=-23.721\pm 0.020$ fm and
$r_{exp}=2.658\pm 0.062$ fm \cite{MW84}
shows that the phase shift is
sufficiently well described by the system of poles
(\ref{1s0poles}). The negative poles correspond to
transformation functions of type $u_j = \exp(a_jx)$.
Positive poles can
correspond either to regular solutions
$u_j = \sinh(a_j x)$  or to
\begin{equation}\label{CHOICE}
u_j = A_j \exp(a_jx)+ B_j \exp(-a_jx)~,
\end{equation}
as implied by (\ref{uj}).
According to the previous section the poles $b_j$ correspond to
regular solutions $v_j=\sinh(b_jx)$.

We can now derive isophase potentials associated with $\delta_6$
by using (\ref{VN}).
The system of poles (\ref{1s0poles}) can correspond
either to a one level potential (deep potential) with $E_0=-a_3^2$ or
to
a shallow potential which has no discrete levels.
The latter
is the desired case and can be obtained when the function $u_3$ is
a $\sinh$ function.
Fig. 1 shows our shallow potential $V_6$ obtained in this way
together with the Reid68 potential
\cite{REID68}

\begin{equation}\label{Reid68}
V_{Reid68}(x)=[-10.463\exp (-0.7x)-1650.6\exp (-2.8x)+
        6484.2\exp (-4.9x)]/(0.7x)~,
\end{equation}
and
the potential of Ref. \cite{BAYE}. The dots represent an
updated version \cite{STOKS} of the Reid soft core potential
\cite{REID68}. In contrast with Ref. \cite{BAYE}
our potential
has a correct short range behaviour $V_6(x)\to
6x^{-2}$ corresponding to the case $\nu = n_+ - n_- = 2$ (the
potential of Ref. \cite{BAYE} has $\nu =1$).
In Fig. 2 we present the tail of the absolute value of various potentials in
a logarithmic scale. One can see that $V_6$ has a smooth behaviour consistent
with the one-meson exchange theory as it is the Reid's potential \cite{REID68},
 in contradistinction to
the SUSY potential from Fig. 3 of Ref. \cite{SPAR97} which displays an
undesired
oscillation around $r=2$~fm\footnote{The potential drawn in
Figs. 1 and 2 under Ref. \cite{SPAR97} represents the output of Eq.
(32) of this reference.}.
Interestingly,  our potential is very close to  the Reid68
potential (\ref{Reid68}) in a large interval up to $5$~fm.

Actually the choice (\ref{CHOICE})
for $u_3$ gives an isospectral and isophase family of $V_6$
potentials with the bound state energy $E_0$ fixed.
In Fig. 3 we display a few deep potentials,
members of this family and isophase
with the shallow $V_6$ potential.
These potentials differ from each other only by the value of the coefficient
$A'_3= A_3/B_3$ in (\ref{CHOICE}).
One can see that by increasing $A'_3$ the potential
well shifts to the right without changing its depth. The bound
state with $E_0= -a_3^2$ follows the movement, being localized
at larger and larger values of $x$.
A similar behavior has been observed  by Bargmann
\cite{BARGMANN} already in 1949
(see also \cite{ZakSuz}).
When $A'_3$
approaches $-1$
the well concentrates near the origin and becomes very narrow and deep.
The ground state becomes well localized
near the origin.
At very short distances a barrier appears
before the well
which grows
when $A'_3\to -1$.
In the limit $A'_3=-1$ the well
becomes infinitely narrow and
i
completely collapses, being replaced by an
infinitely high barrier which becomes
a repulsive core,
as for a shallow potential.

One can also construct a deep potential
isophase with the shallow potential $V_6$ but having a discrete
level at a desired place. First one has to delete the bound
state $E_0=-a_3^2$ of $V_6$ following the procedure described in Sec III.
Next one has to use a
$2$-parameter family of
isophase potentials (Subsection III C).
This is possible by
adding two new
transformation functions, which gives a total of $8$ functions.
The new functions $u_4$ and $v_4$ must have $a_4=-b_4$
($a_4 < 0$) in order to
give $\delta_8(k)= \delta_6(k)$. It is enough to take $u_4 = \exp(b_4x)$
and calculate the Wronskian of the added pair from  (\ref{W2}).
When $|a_4|=a_3$
the potential behaves like a deep $V_6$ potential
with the energy $E=-a_3^2$. By varying $a_4$
one can change the depth of the well.
The family of $V_8$ potentials
has in fact two parameters: the energy $E=-a_4^2$ and the parameter $c$
of (\ref{W2}).
One can select a potential from this
family by fixing these two parameters. Here we look for a deep
potential which describes the nucleon-nucleon interaction
according to microscopic theories based on the quark structure of nucleons
(see e. g. Refs. \cite{NEUD75} or \cite{KUKULIN}).
We wish to obtain a potential
close
to the deep potential $V_K$
of Ref. \cite{KUKULIN},
given by the formula
\begin{equation}\label{KUK}
V_K=-1106.21\exp (-1.6x^2)-
10.464\exp (-0.7x)/(0.7x)[1-\exp (-3x)]~,
\end{equation}
where $x$ is given in fm. We varied $a_4$ and $c$ such as to be
as close as possible to (\ref{KUK}). From a least square fit
we obtained $c= - 0.155$ fm and
$a_4=-3.7944$ fm$^{-1}$ which gives $E$ = 596.42 MeV.
In this way we obtained the potential $V_8$
which is compared to $V_K$ of (\ref{KUK}) in
Fig. 4.
One can see that the two potentials
are quite close to each other. The bound state of $V_K$ is 442.MeV
which is not much different from our value.
The form of $V_K$ was chosen on qualitative grounds related to the
approximations used in microscopic studies of the nucleon-nucleon
interaction \cite{NEUD75}. The potential $V_8$ of Fig. 4 is as legitimate
as $V_K$
inasmuch as it correctly describes the experimental phase shift.
The difference between the two potentials is partly due the difference
in the fitted
phase shifts and the $E_{lab}$ interval chosen for the fit in each case.
If
instead of fitting the phase shift to the experimental values
we were fitting it to
the phase shift corresponding to the potential (\ref{KUK}),
the difference would have been smaller.

\section{CONCLUSIONS}

In this study we propose an alternative method to derive
families of phase equivalent (isophase) local potentials
based on the application of an $N$th order Darboux transformation
or an equivalent chain of $N$ first order transformations.
In practical applications this method
is
simpler and more efficient than the conventional
SUSY approach of Refs. \cite{SUKUMAR,BAYE,SB97,SPAR97,SPAR2000,SBL}.
First one has to look for imaginary poles of the
S-matrix, i.e. one has to find the parameters $a_j$ and
$b_j$ of the phase shift  $\delta_N(k)$ of (\ref{deltaN})
which fit a particular experimental phase shift.
One can start with the lowest order $N$ = 2 in
(\ref{deltaN})      and increase it until a good fit
of the phase shift is achieved.
The second step consists in finding an appropriate
mapping between the $S$-matrix poles found in this way and the system of
functions (\ref{uv}) which give a nodeless Wronskian.
In terms of these functions the formula (\ref{VN})
gives phase
equivalent potentials, among which one is shallow
and the others form families of deep potentials from which one can
select a particular one for a specific physical problem.
We have applied our method to derive potentials describing the
neutron-proton experimental phase shift for the $^1S_0$
partial wave. We obtained a shallow
potential very close to the Reid's soft core potential and its
deep phase equivalent partner is close to potentials resulting
from microscopic derivations of the nucleon-nucleon interaction.
In order to obtain phase equivalent potentials describing higher partial waves
$l$-changing Darboux chains of transformations are necessary.
This is the scope of a subsequent study.

\appendix
\section{}

Here we prove the formula (\ref{MNnu}). The proof is split into two steps:
1) Taking an initial potential with $\nu =\nu _0=0$
we prove (\ref{MNnu}) for this case.
2) We introduce a number of
regular solutions to the previous case
which corresponds to $\nu >0$.
Finally we comment on
singular initial potentials.

Let us consider a chain of  $N=2n$ Darboux transformations
for a potential $V_0(x)$ with $\nu =\nu_0=0$ and
transformation functions given by (\ref{uv}) with $\nu =0$.
Since all $a$'s and $b$'s are different from each other this
chain is equivalent to an $N$th order transformation.
The
transformed potential $V_N$ is given by (\ref{VN}) and the eigenfunctions
$\psi_N$
by (\ref{psiN}) or (\ref{psiNj}). One can also find the Jost
solution of  $h_N$ by applying the same operator $L^{(N)}$ to
the Jost solution of $h_0$. The result is the function $\tilde
f_N(x,k)=L^{(N)}f_0(x,k)$, which like $\psi_N$ from
(\ref{psiN}), is proportional to the ratio of two determinants.
The long-distance
asymptotic value of $\tilde f_N(x,k)$ is
proportional to the asymptotic value of the Jost solution.
Therefore using the usual definition of the Jost function as
the value of the Jost solution at the origin
\cite{FADDEEV,Chadan-Sabatier} one has
\begin{equation}\label{RATIO}
F_N(k)=
\frac{\tilde f_N(0,k)}
{[e^{-ikx}{\tilde f_N(x ,k)]_{\,x\to\, \infty\,}}^{\vphantom{A}}}
\end{equation}
The definition (\ref{psiN})
and our special
choice of transformation functions (\ref{uv}) implies that
\begin{equation}\label{JOST}
\tilde f_N(x,k) = W(u_1,v_1,\ldots ,u_n,v_n,f _0(x,k))\,
W^{-1}(u_1,v_1,\ldots ,u_n,v_n) \equiv W_f \, W^{-1}_N
\end{equation}
Therefore both for $x = 0$ and $x\to \infty $
we have to calculate the ratio of two determinants
of orders $N + 1$ and $N$ respectively, but of a similar structure.

First let us consider the function $\tilde f_N(x,k)$ at $x=0$.
Using the definition $f_0(0,k) = F_0 (k)$
the determinant appearing in the numerator of (\ref{JOST}) becomes
$$
W_f =
\left|
\begin{array}{llllllll}
C_1       &  0      & C_2    &0       &\ldots &C_n     &0       &F_0(k)   \\
c_1   & D_1         &c_2    &D_2     &\ldots &c_n      &D_n&\tilde F(k)\\
C_1a_1^2  &  0      &C_2a_2^2 &0      &\ldots &C_na_n^2&0       &F_0(k)(ik)^2 \\
c_1a_1^2  & D_1b_1^2&c_2a_2^2 &D_2b_2^2&\ldots &c_na_n^2 &D_nb_n^2&\tilde F(k)(ik)^2 \\
\vdots    & \vdots  &\vdots  &\vdots  &\ddots &\vdots  &\vdots  &\vdots \\
C_1a_1^{2n-2}&0     &C_2a_2^{2n-2}&0      &\ldots &C_na_n^{2n-2}&0   &F_0(k) (ik)^{2n-2}\\
c_1a_1^{2n-2}&D_1b_1^{2n-2}&c_2a_2^{2n-2}&D_2b_2^{2n-2}&\ldots &c_na_n^{2n-2} &D_nb_n^{2n-2}&\tilde F(k) (ik)^{2n-2}\\
C_1a_1^{2n}&0     &C_2a_2^{2n}&0      &\ldots &C_na_n^{2n}&0   &F_0(k)(ik)^{2n}
\end{array}
\right|
$$
$$
=C_1\ldots C_nD_1\ldots D_nF_0(k)
\left|
\begin{array}{lllll}
1&1&\ldots &1&1\\
a_1^2&a_2^2&\ldots &a_n^2&(ik)^2\\
\vdots &\vdots &\ddots &\vdots &\vdots\\
a_1^{2n-2}&a_2^{2n-2}&\ldots &a_n^{2n-2}&(ik)^{2n-2}\\
a_1^{2n}&a_2^{2n}&\ldots &a_n^{2n}&(ik)^{2n}\\
\end{array}
\right|
\times
\left|
\begin{array}{lll}
1&\ldots &1\\
b_1^2&\ldots &b_n^2\\
\vdots  &\ddots &\vdots \\
b_1^{2n-2}&\ldots &b_n^{2n-2}
\end{array}
\right|
$$
\begin{equation}\label{Det}
=-F_0(k)\prod\limits_{j\,=1}^n(C_jD_j)
\prod\limits_{m>j}^n(a_j^2-a_m^2)(b_j^2-b_m^2)
\prod\limits_{j\,=1}^n(a_j^2+k^2)
\hphantom{MMMMMMMMM}
\end{equation}
Here $C_j=u_j(0)$, $c_j=u_j'(0)$, $D_j=v'_j(0)$,
$\tilde F(k)=f'_0(0,k)$.
Note that the derivative of the Jost function, $\tilde F(k)$,
does not appear in the final expression for $W_f$.
The above result has been obtained
by first using
the following
determinant property
\begin{equation}\label{detM}
\left|
\begin{array}{cc}
M_{p\times p}&0\\
\cdots  & \widetilde M_{q\times q}
\end{array}
\right|
=\det M_{p\times p}\det \widetilde M_{q\times q}
\end{equation}
and next by expanding the two Vandermonde determinants
(see e.g. \cite{LANKASTER})
from the middle line.

The determinant $W_N$
which appears in the
denominator of (\ref{JOST}) is of even order. It is obtained
from the above determinant  by the removal of the last row and column.
Its final form is similar to the one above.
It has opposite sign and
the factors depending on $k$ are absent.
Finally one finds
\begin{equation}\label{F0}
\tilde f_N(0,k)=(-1)^nF_0(k)\prod\limits_{j\,=1}^n(k^2+a_j^2)
\end{equation}

Let us now
consider the denominator of (\ref{RATIO}), i. e.
${\tilde f_N(x ,k)}^{\vphantom{A}}$ for
$x\to \infty $.
In this limit all functions
entering the Wronskians of (\ref{JOST}) tend to
exponentials which
cancel out with each other
except for $\exp (ikx)$ which cancels out
with the exponential
present in (\ref{RATIO}). The remaining factor
is a ratio of two Vandermonde determinants
of orders $N+1$ and $N$ respectively
containing powers of $a_j$ and $b_j$.
The determinant in the numerator differs from that in the denominator
by the presence of a column containing powers of
$ik$ and an additional row composed of $(N+1)$th power of
$a_1,\ b_1,$ \ldots  $a_n,\ b_n,\ ik$.
Using once more the expression
of Vandermonde's determinant
one obtains
 \begin{equation}\label{asinf}
\tilde f_N(x,k)|_{x\to \infty}\longrightarrow
i^N\prod\limits_{j\,=1}^n(k+ia_j)(k+ib_j)
 \end{equation}

Introducing (\ref{F0}) and (\ref{asinf}) in (\ref{RATIO})
one gets the following relation between the initial
and final Jost functions
\begin{equation}\label{MNk}
F_N(k)=F_0(k)\prod\limits_{j\,=1}^n\frac{k-ia_j}{k+ib_j}~.
\end{equation}
Moreover, since the Wronskian $W_N$ is different from zero at $x=0$ the
singularity strength parameter of $V_N$ is zero and it is
regular at the origin.

Consider now the case where the first $\nu $ functions of (\ref{uv})
belong to the
regular family and calculate the singularity strength
 of
$V_N$.
Note that
the system (\ref{uv}) can be divided into two subsystems,
one containing an equal number of singular and regular
functions, and the other composed of regular functions only. Accordingly
the $N$th order transformation can be decomposed into a superposition of
two transformations. The first transformation of order $N-\nu $
which is based on functions paired by their regular property will not
affect the regularity of the potential at the origin,
whereas the second one of order $\nu $ composed of regular
functions only will increase the singularity parameter of the potential by one
unit for each transformation function (see e.g. \cite{BAYE}).
So, the singularity strength parameter of $V_N(x)$ is equal to $\nu $.

In the following we have to use the definition of the Jost
function
\cite{FADDEEV,Chadan-Sabatier}
\begin{equation}\label{Jostfunction}
F(k)=
\lim\limits_{x\to \,0}
\frac{(kx)^{\nu}}{i^{\,\nu}(2\nu -1)!!}f(x,k)
\end{equation}
We give detailed calculations for the particular case
$\nu =2$ from which a general formula can easily be derived.
In this case
both $W_f=W(v_1,v_2,u_3,v_3,\ldots ,u_n,v_n,f_0)$  and
$W_{N}=W(v_1,v_2,u_3,v_3,\ldots ,u_n,v_n)$  are equal to zero at $x=0$.
Since $f_0(0,k)\ne 0$ only one function (either $v_1$ or $v_2$) in $W_f$
is an unpaired regular solution and hence the zero of $W_f(0)$ is
of the first degree.
For $\nu =2$ the second logarithmic
derivative of $W_N(x)$ at $x=0$ is of order $\nu (\nu +1)=6$. We
conclude that the zero of $W_N(0)$ is of the third degree.
 To evaluate the  indeterminate form
defining the Jost function
(see (\ref{Jostfunction}) for $\nu =2$ and $f(x,k)=f_N(x,k)$)
 one can use  L'H\^ospital's rule to get
 $(kx)^2~W_f/W_N=k^2~W_f/x*x^3/W_N\to 3!~k^2~W_f'(0)/W_N'''(0)$
 when $x\to 0$,
because the first two derivatives of $W_N(0)$ are zero.
Since the derivative of an $N$th order Wronskian
$W(f_1,\ldots ,f_N)$
 is a determinant obtained
from the same Wronskian by the replacement of the last row
composed of the derivatives $f_j^{(N-1)}$ by the derivatives
$f_j^{(N)}$
one can use
the property (\ref{detM}) to factorize  $W_f'(0)$ into a product
of two Vandermonde determinants $W_{V1}$ and $W_{V2}$
 of orders $n$ and $n+1$ respectively.
When calculating
the third derivative of $W_N$
at $x=0$
one can see that it
is equal to a sum of two equal determinants
which can be
factorized according to (\ref{detM}) in a product of two
Vandermonde determinants
$W_{V3}$ and $W_{V4}$ of orders $n-1$ and $n+1$ respectively.
We do not give here the explicit form of these determinants.
The factor
$2$ coming from two equal Vandermonde determinants
will cancel out with the corresponding factor coming from $3!$
and the factor $3$ from this factorial
will cancel out with $(2\nu -1)!!$ for $\nu =2$,
the latter being present in the definition of the Jost function
(\ref{Jostfunction}).
Hence, $(kx)^2\,W_f/W_N\to k^2W_{V1}W_{V2}/(W_{V3}W_{V4})$.
When evaluating the corresponding Vandermonde determinants one can
see that the factors independent
of $k$ are exactly the same
 both in the numerator and the denominator.
So, one gets the following Jost function for $V_N(x)$
\begin{equation}\label{MNk1}
F_N(k)=
F_0(k)\frac{k^2}{(k+ib_1)(k+ib_2)}
\prod\limits_{j\,=\,3}^n\frac{k-ia_j}{k+ib_j}
\end{equation}
It is clear from here that every additional regular function in
the system (\ref{uv}) with the parameter $b_j$ produces a factor
$k/(k+ib_j)$ in the Jost function which corresponds exactly to
the formula (\ref{MNnu}). Moreover, a regular at the origin
potential admits two kinds of transformations, the ones which do
not affect its singularity ($\nu =0$ in (\ref{uv})) and others
which increase its singularity ($\nu >0$).
Actually if the initial potential has
the singularity parameter $\nu_0\ne 0$,  three types
of transformations are possible. In addition to two the previous types
of transformations
existing for the case $\nu_0=0$ transformations which
decrease the singularity parameter are also possible. All these
possibilities become clear if one takes into consideration the
fact that all transformations are invertible
(this means that from an intermediate element of a chain one can move in both
directions)
 and an intermediate
member of a previously considered chain is just a potential with a
singularity strength different from zero.

\begin{center}
{\bf ACKNOWLEDGMENTS}.
\end{center}

We are most grateful to Mart Rentmeester for providing
details about the Reid93 potential. Illuminating discussions
with Daniel Baye and Francis Michel are gratefully acknowledged.
One of us (B.F.S) is grateful to the
Theoretical Fundamental Physics Laboratory of the University
of Li\`ege for warm hospitality during
autumn 2001 when this work has been started.
He also acknowledges a partial support from RFBR.



\begin{table}\label{I}\caption{The $^1S_0$ experimental phase
shift from Ref. \protect\cite{STOKS}
together with the theoretical value $\delta_6 $.
$E_{lab}$ is related to the momentum $k (fm^{-1})$ by $k=
\sqrt \frac{m_n E_{lab}}{2 \hbar^2 c^2}$ where we took $m_n$ = 940 MeV
 and $\hbar c = 197.33$ MeV~fm.}
\vspace{.5em}
\renewcommand{\arraystretch}{0.8}
\begin{tabular}{rrr}
$E_{lab}$ (MeV)  & $\delta_{exp}(deg)$ & $\delta_6 (deg)\vphantom{\int_A}$ \\
\tableline
$\vphantom{A^{A^A}}$
  0.100  &  38.43000 & 38.422 \\
  4.000  &  64.28440 & 64.362 \\
  9.000  &  60.72287 & 60.717 \\
 17.000  &  55.34120 & 55.253 \\
 28.000  &  49.39220 & 49.291 \\
 42.000  &  43.32663 & 43.270 \\
 56.000  &  38.29709 & 38.291 \\
 70.000  &  33.95623 & 33.988 \\
 84.000  &  30.11287 & 30.167 \\
 98.000  &  26.64896 & 26.711 \\
112.000  &  23.48572 & 23.546 \\
126.000  &  20.56759 & 20.620 \\
140.000  &  17.85371 & 17.893 \\
154.000  &  15.31309 & 15.337 \\
168.000  &  12.92158 & 12.930 \\
182.000  &  10.66001 & 10.654 \\
196.000  &   8.51284 & 8.4939 \\
210.000  &   6.46736 & 6.4377 \\
224.000  &   4.51295 & 4.4752 \\
238.000  &   2.64074 & 2.5979 \\
252.000  &    .84314 & 0.7984 \\
266.000  &   -.88631 &-0.9299 \\
280.000  &  -2.55320 &-2.5924 \\
294.000  &  -4.16240 &-4.1942 \\
308.000  &  -5.71816 &-5.7397 \\
322.000  &  -7.22423 &-7.2326 \\
336.000  &  -8.68393 &-8.6765 \\
350.000  & -10.10022 &-10.074 \\

\end{tabular}
\end{table}

\begin{figure}
\centering \epsfig{file=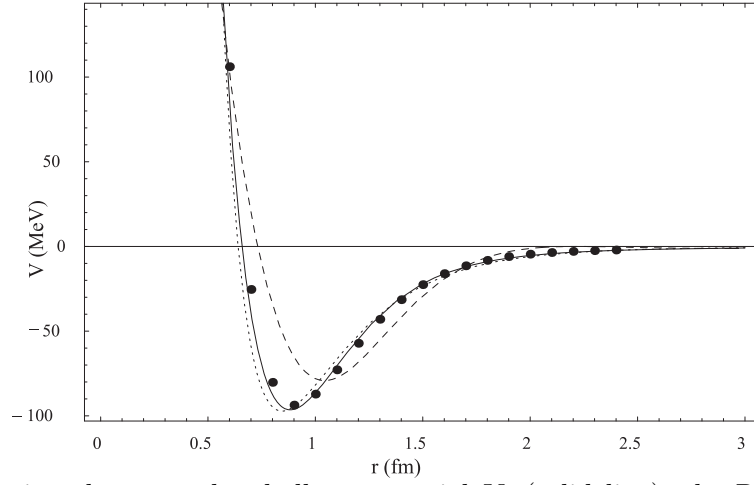, width=10cm}
\caption{\small Comparison between the shallow potential $V_6$ (solid line),
the Reid68 potential \protect\cite{REID68} (dotted line), the potential of Ref.
\protect\cite{SPAR97}
(dashed line) and the Reid93 potential \protect\cite{STOKS} (dots)
represented as a function of $x \equiv r$.}
\end{figure}

\begin{figure}
\centering \epsfig{file=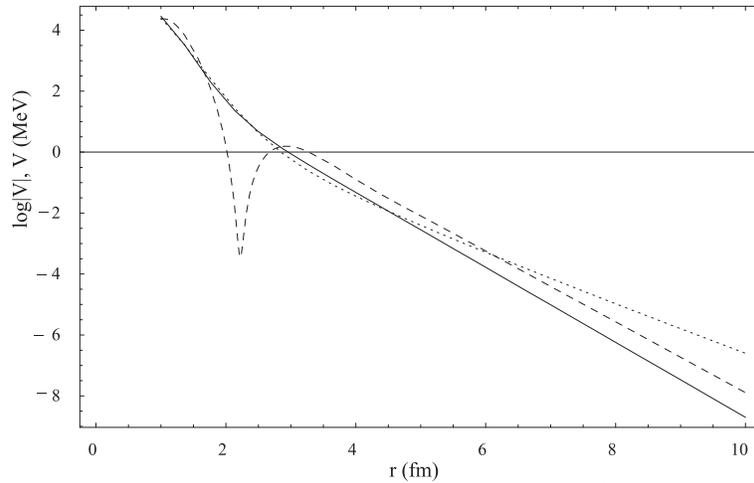, width=10cm}
\caption{\small  The asymptotic behaviour
of the absolute value of
$V_6$ (solid line), the
Reid68 potential \protect\cite{REID68} (dotted line)
and the potential of Ref.
\protect\cite{SPAR97}
(dashed line) in (natural) logarithmic scale.}
\end{figure}
\vspace{5em}

\begin{figure}
\centering \epsfig{file=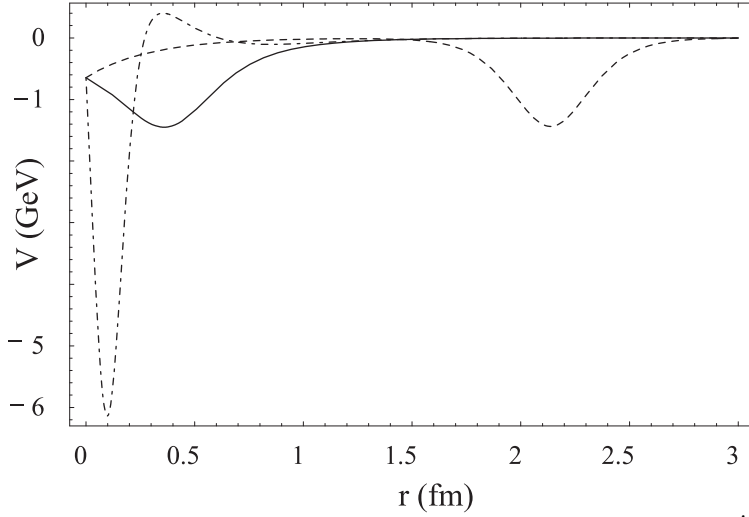, width=10cm}
\caption{\small Family of isospectral potentials isophase with $V_6$.
Solid line - $A'_3=0$; Dashed line - $A'_3=10^6$;
dash-dotted line --- $A'_3=-0.95$.
For the latter case appearance of a barrier is clearly visible.
}

\end{figure}

\begin{figure}
\centering \epsfig{file=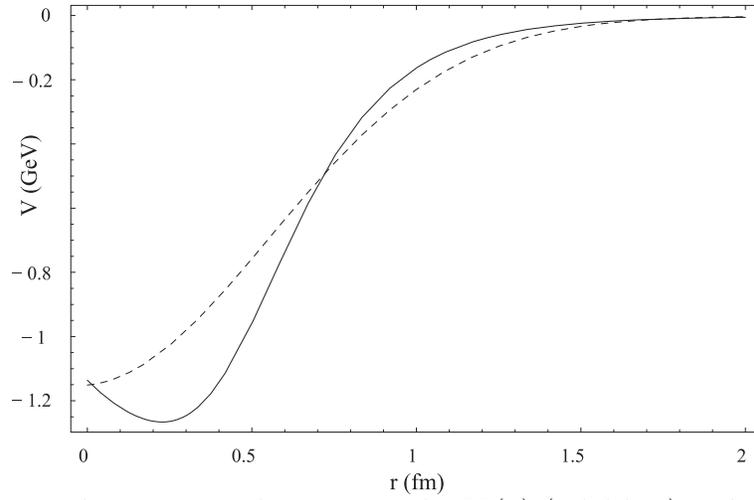, width=10cm}
\caption{\small Comparison between two deep potentials:  $V_8(x)$
 (solid line) and the potential of Ref. \protect\cite{KUKULIN} (dashed line).}
\end{figure}

\end{document}